\documentclass[%
 aip,
 amsmath,amssymb,
preprint,%
]{revtex4-1}

\usepackage{graphicx}
\usepackage{dcolumn}
\usepackage{bm}

\usepackage[utf8]{inputenc}
\usepackage[T1]{fontenc}
\usepackage{mathptmx}
\usepackage{etoolbox}

\makeatletter
\def\@email#1#2{%
 \endgroup
 \patchcmd{\titleblock@produce}
  {\frontmatter@RRAPformat}
  {\frontmatter@RRAPformat{\produce@RRAP{*#1\href{mailto:#2}{#2}}}\frontmatter@RRAPformat}
  {}{}
}%
\makeatother
\begin{document}

\preprint{AIP/123-QED}

\title[Shot-to-shot acquisition ultrafast electron diffraction]{Shot-to-shot acquisition ultrafast electron diffraction}
\author{Rémi Claude}
\author{Michele Puppin}%
\author{Bruce Weaver}%
\author{Paolo Usai}%
\author{Thomas LaGrange}%
\author{Fabrizio Carbone}
\email{fabrizio.carbone@epfl.ch}

\affiliation{Laboratory of Ultrafast Microscopy and Electron scattering, Ecole Polytechnique Fédérale de Lausanne}

\date{\today}

\begin{abstract}
We demonstrate a novel shot-to-shot acquisition method for optical pump - keV electron energy probe in ultrafast scattering experiments. We integrate a phase-locked acquisition scheme at a repetition rate of 20kHz in a conventional ultrafast electron diffraction (UED) setup. We proceed to a full characterization of the noise level in different configurations and for realistic scenarios. The shot-to-shot acquisition improves the signal-to-noise ratio (SNR) by one order of magnitude and can be readily implemented in other high-repetition-rate electron diffraction and spectroscopy setups.
\end{abstract}

\maketitle

\section{\label{sec:level1}Introduction}

Ultrafast electron diffraction, microscopy, and energy loss spectroscopy are widely used to investigate the properties of materials out-of-equilibrium using pump-probe methods. Transient signals recorded by these techniques correspond to a variation in the number of scattered electrons at specific positions in the momentum, real, or energy space, which enables studies in the temporal domain of lattice dynamics \cite{carbone_structural_2008, Chatelain2014}, magnetic ordering \cite{ULTEM, berruto_laser-induced_2018, rubiano_da_silva_nanoscale_2018} and demagnetization \cite{tauchert_polarized_2022},  plasmon dynamics \cite{piazza_simultaneous_2015}, charge density waves coherent response and melting \cite{cheng_ultrafast_2024}, and to resolve changes in the phonon population with momentum resolution\cite{rene_de_cotret_time-_2019, stern_mapping_2018, waldecker_momentum-resolved_2017}. The amplitude of transient signals increases with the photoexcitation energy density and is typically on the order of a few percent for light excitation in the range of mJ/cm$^2$. 
Advanced materials science focuses on broken-symmetry ground states, where a delicate microscopic balance of electron, spin, lattice, and orbital degrees of freedom is achieved. In electron diffraction, elastically scattered electrons provide measurements of lattice order, while inelastically scattered electrons contain crucial information on the momentum-dependent interaction between collective modes \cite{van_der_veen_ultrafast_2015, stern_mapping_2018}. For instance, phonon populations alter thermal diffuse scattering, with a typical total cross-section of $10^{-4}$. Other quasiparticles, such as magnons, are also present in the diffuse inelastic scattering but represent only $10^{-6}$ of the total electron flux \cite{lyon_theory_2021, kepaptsoglou_magnon_2025}. Time-resolved measurements of such inelastic signals present a significant challenge for current experimental setups, which must improve detection sensitivity to capture small photo-induced variations, particularly in weak perturbation regimes that maintain the integrity of the ground state.

To improve the signal-to-noise ratio (SNR), a first possibility is to develop a brighter electron source ($>10^5$ electrons/pulse). However, this approach introduces space-charge-induced temporal broadening of the electron pulse, which significantly degrades the time resolution \cite{siwick_ultrafast_2002, wang_measurement_2009}. Temporal compression of the electron bunch can be achieved using a synchronized radio frequency cavity \cite{mancini_design_2012, chatelain_ultrafast_2012}, or a terahertz optical pulse \cite{ehberger_terahertz_2019}. However, special care must be taken to avoid jitter and long-term drifts \cite{otto_solving_2017}. A second approach is to minimize the noise of the electron source while working at low electron flux ($<10^4$ electrons/pulse). To mitigate space charge broadening, the electron source is placed as close as possible to the sample \cite{waldecker_compact_2015}. Combined with a high repetition rate laser source and fast acquisition system\cite{freelon_design_2023, aidelsburger_single-electron_2010}, the SNR is increased while maintaining good beam coherence. A possible limitation of this method is the sample re-equilibration time, which can become greater than the inter-pulse separation ($<1\mu$s) \cite{vidas_does_2020}. \\

In this work, we present an ultrafast electron scattering apparatus based on a 20 kHz repetition rate source operating at low flux, with fewer than one thousand electrons per pulse. In this limit, the quality of the acquisition system is essential to improve the SNR. We employ a direct electron detector\footnote{Quadro, Dectris}, which records high SNR electron diffraction patterns \cite{mcmullan_chapter_2016} thanks to a combination of high detective quantum efficiency (DQE) (0.9 for direct detection, as compared to 0.2 for a typical charge-coupled device (CCD)) \cite{levin_direct_2021} and low acquisition noise. Another advantage of such a direct detection system is its capability to acquire data at a high rate, synchronized with the laser repetition rate \cite{duncan_multi-scale_2023}. This enables the collection of pump-probe data by chopping the pump pulse train at 10 kHz and normalizing the data using unpumped acquisition frames. This method, well-established in ultrafast optical spectroscopy due to the availability of fast light detectors, is widely used to achieve shot-noise-limited data \cite{lang_photometrics_2018}. In electron scattering experiments, where the number of events per pixel and per pulse is low, the method remains viable due to the absence of the detector’s dark count contribution. We demonstrate pump-probe data with a noise level reaching $10^{-4}$ in a few hours of acquisition time, with sub-picosecond time resolution. \\
The paper is organized as follows. In section \ref{sec:part2}, we describe the setup, the acquisition system, and the working principles of the synchronization to achieve shot-to-shot acquisition. Section \ref{sec:3} presents the characterization of the noise level. In section \ref{sec:4}, we will present UED data acquired in a realistic scenario by photo-exciting graphite samples with an 800~nm pump. 

\section{\label{sec:part2} Design of the shot-to-shot UED}
\begin{figure*}
\includegraphics[width=\linewidth]{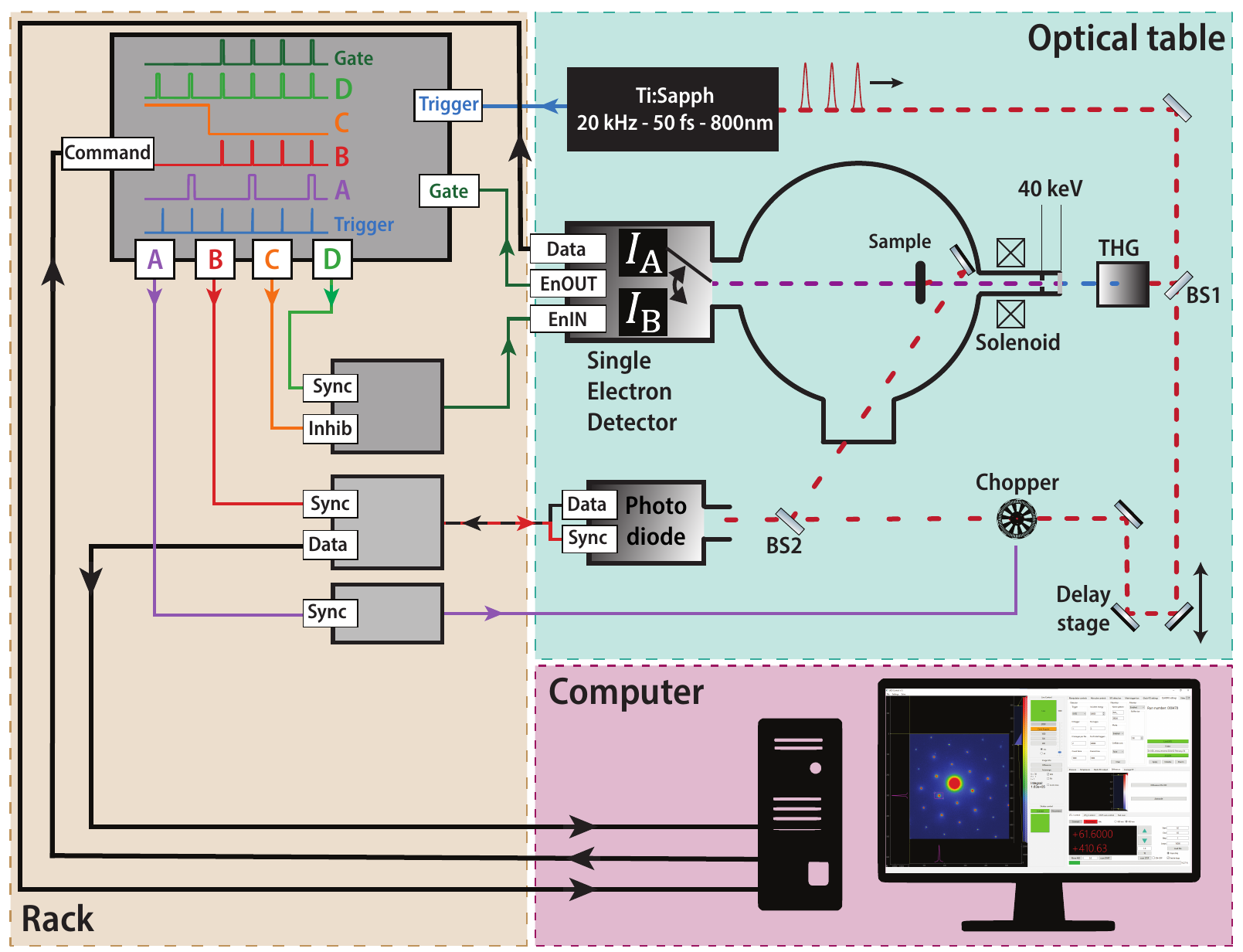}
\caption{\label{fig:UED} Scheme of the UED setup in EPFL composed of the optical table, the rack, and the computer. The optical table supports the pulsed laser, the optical beam path divided into the pump and the probe arms, the vacuum sample chamber, the photodiode, and the electron detectors. The rack accommodates the synchronization system, which is phase-locked to the laser source at 20 kHz. The data gathered by the electron and photodiode detectors are sent to the computer to be sorted between pumped and unpumped diffraction patterns. }
\end{figure*}

\subsection{\label{sec:part2a}Optical setup and electron diffraction beamline}
The ultrafast electron diffraction (UED) setup and the data acquisition scheme are schematized in Fig. \ref{fig:UED}. The current setup is a modification of the setup detailed in Ref. \cite{mancini_design_2012}.
A cryogenically cooled Ti:Sapphire laser\footnote{Wyvern, KM Labs} provides 60 fs pulses at a central wavelength of 780 nm at a repetition rate of 20 kHz with an output power higher than 10 W (pulse energy > 0.5 mJ). To avoid long-term drifts, the laser beam pointing is stabilized by a home-built stabilization system based on a pair of piezoelectric-actuated mirrors and CMOS cameras. The beam is divided into two arms with a beam splitter (BS1). 
The pump beam goes through a mechanical delay line to vary the delay between the probe pulse and the pump pulse. It then passed through a mechanical chopper, synchronized with the laser source, in order to halve the pulse repetition rate to 10 kHz. The last mirror (BS2) sends the pump into the chamber and is used to find spatial overlap between the pump beam and the electron probe. BS2 also acts as a beam splitter, transmitting a small fraction of the pump towards a photodiode.  \\
In the probe arm, third harmonic generation (THG) is used to generate a UV light beam (260 nm) which back-illuminates a silver-coated sapphire photo-cathode. The UV probe pulse energy exceeds the work function of the silver ($P_{\text{UV}} = 4.76 \text{ eV} > W_{\text{Ag}} = 4.26$ eV), thus generating a pulse of electrons by single-photon photoemission. In Appendix \ref{App:A}, we find the optimal condition for the photoemission when the measured quantum efficiency is maximum at $10^{-8}$ electrons/photon. It corresponds to an acquired flux between 1 and 1000 electrons/pulse photoemitted at UV powers from 1 $\mu$W and 1 mW, respectively. These pulses are then accelerated to 40 keV over a distance of 3 mm, corresponding to an accelerating electric field of 13.3 MV/m. The off-axis photoemitted electrons are filtered out using a 100 $\mu$m pinhole after the anode, providing the low beam emittance needed for coherent electron scattering with high reciprocal resolution. A solenoidal magnetic lens with approximately 3000 windings, each carrying a current of 0.62 A, focuses the electron beam onto the detector. Operating at 17.3 W, the lens achieves a focal length of 58 mm. At the sample holder position, 15 cm away from the electron source, the electron beam waist has a diameter of approximately 200 $\mu$m. For an electron flux of nearly 1000 electrons/pulse on the detector, we obtain a time resolution of around 1 ps. \\
In transmission geometry, the sample holder hosts five standard 3 mm TEM grids. The holder is mounted on a 4-axis open-cycle cryo-manipulator that can achieve a temperature of 4.2K with liquid helium cooling. The UED setup can also work in reflection geometry to probe the surface lattice dynamics with five axes of freedom using another sample holder \cite{pennacchio_design_2017}. \\
The vacuum chamber has a base pressure of $10^{-8}$ mbar and accommodates a mirror that reflects the pump beams onto the sample at an angle of a few degrees off from the electron beam. \\
The electron beam travels a distance of 57 cm after the sample position. At this distance and with a chip detection size of $38.4\times 38.4$ mm$^2$, a reciprocal space area of $16 \times 16$ \AA$^{-2}$ is covered at 40 keV. 

\subsection{\label{sec:part2b}High speed lock-in direct electron detection}

The direct electron sensor is Si-based with $512\times512$ hybrid-pixels with a DQE of 0.9 at 100 keV and can acquire electrons from 30 keV to 200 keV, without readout noise. However, after an electron collides with a pixel, it remains triggered for 50 ns. Therefore, the detector cannot acquire more than one electron per pulse per pixel, which reduces the DQE for a high pulsed electron flux, as explained in Appendix \ref{App:B}. The fast electronics allow an acquisition rate of 2250 frames per second and up to 18000 frames per second for a reduced region of interest (ROI). Each pixel has two 16-bit counter chips to reach a bit depth of 32 bits. Importantly for this work, it is also possible to switch between these 16-bit counters within 200 ns, sorting electron events independently in each of the two counters. The acquisition window is externally controlled by a time-to-live signal at a rate of 20 kHz, locked to the laser repetition rate. We note that rates exceeding 1 MHz are, in principle, possible by this method. After a given number of acquisition windows (typically 4$\cdot 10^4 \equiv 2$ sec), the total number of counts acquired for the two counters of each pixel is transferred as two full-frame images, labeled $I_{\text{A}}$ and $I_{\text{B}}$ in the figure. In this way, the effective repetition rate of the acquisition is greatly improved while minimizing the data transfer.

We use this detector feature to acquire ultrafast electron diffraction by chopping the pump pulses at half the laser repetition rate (10 kHz). The two output images correspond to the excited, $I_{\text{on}}$, and non-excited, $I_{\text{off}}$, diffraction patterns of the studied sample. Due to the high shot-to-shot correlation of the laser source, this considerably reduces the noise in the pump-probe signal, as will be discussed in Sec. \ref{sec:3}. We now describe the synchronization system in more detail. 

\subsection{\label{sec:part2c}Synchronization system}

To capture the single shots, we synchronize the detection with the time of arrival of the electrons to the laser source repetition.\\
As shown in the left panel of Fig. \ref{fig:UED}, a homemade delay generator constantly receives the 20 kHz laser signal from the laser source. This trigger is relayed into four independently delayed channels (A-D). Channel A is frequency-divided to 10 kHz and transmitted to the chopper controller. 
Channel B triggers a fast photodiode digitizer \footnote{GlazPD, Synertronic design}. This signal in this channel is gated by the main detector (EnOUT), which is a replica of the one received (EnIN). In this way, the Quadro detector operates as the master and the photodiode as the slave, ensuring that the same pulses originating from the laser are detected. The electron detector acquisition window is controlled by channel D, with a 20 kHz repetition rate of a 25 $\mu$s wide square signal.
The acquisition window from channel D is inhibited by channel C, acting like a switch controlled by the main computer to start the acquisition. 
This synchronization system enables robust acquisition of every electron bunch at 20 kHz and can be readily adapted to other repetition rates. However, operation at higher frequencies would require a different chopping mechanism. From the main computer, the number of exposition windows as well as other detector settings can be monitored. \\
The main computer accesses the two 2D datasets acquired by the electron detector, $I_{\text{A}}$ and $I_{\text{B}}$, which can be sorted into $I_{\text{on}}$ and $I_{\text{off}}$, thanks to the photodiode readout. A Python-based user interface \cite{github_Usai} displays this data and the difference between $I_{\text{on}}-I_{\text{off}}$, allowing for a fast and accurate time and spatial overlap.  

\section{\label{sec:3}Characterization}
Given the low shot-to-shot fluctuation of the laser source, this acquisition method greatly improves the SNR of ultrafast experiments. We characterize this enhancement by disentangling the noise sources for conventional and shot-to-shot acquisition techniques with and without the photoexcitation pulse. 

\subsection{\label{subsec:3.a} Noise level without pump}

\begin{figure}
\includegraphics[width=\linewidth]{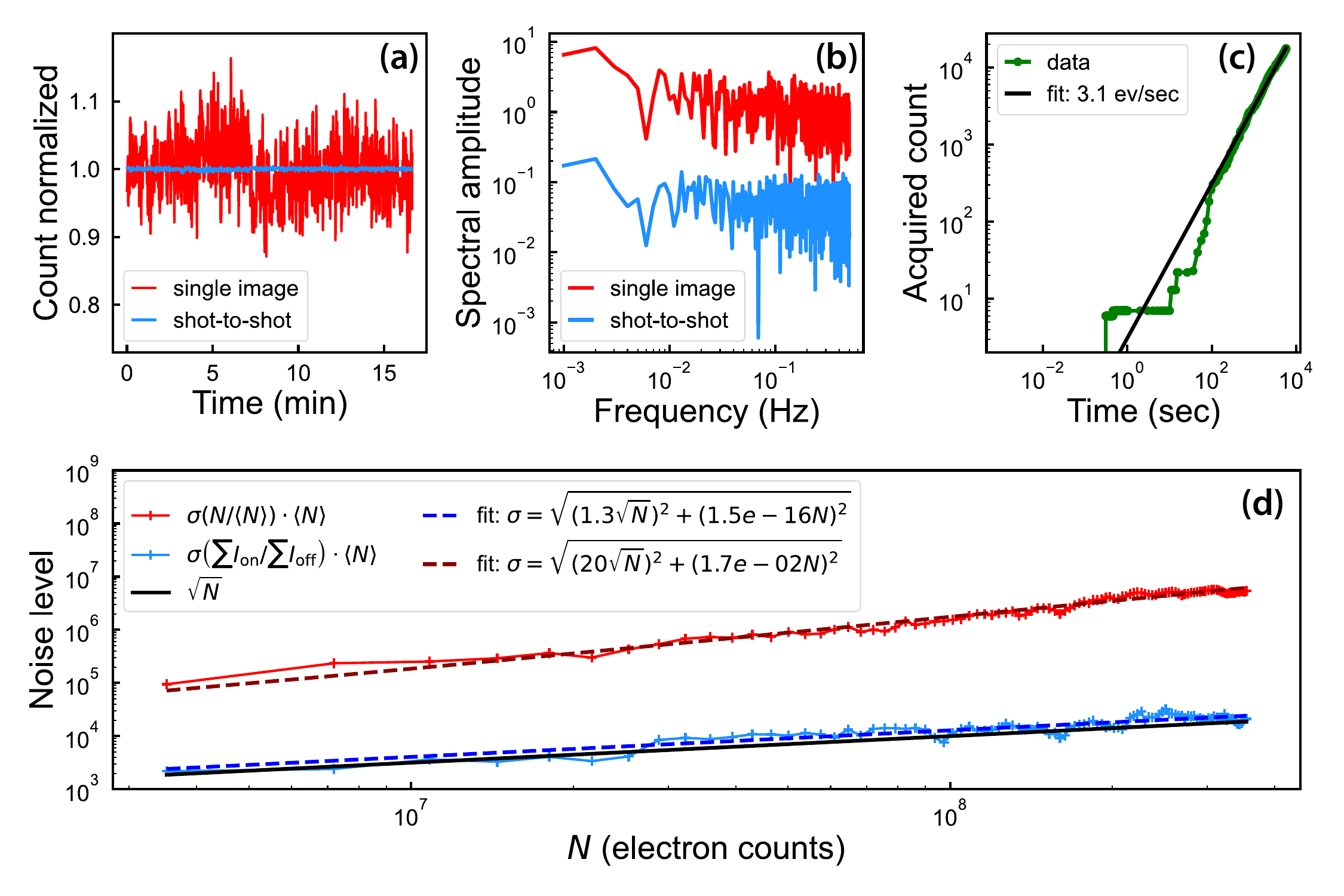}
\caption{\label{fig:fluctuation} a) Fluctuation of the electron diffraction pattern of graphite sample in the absence of a pump as a function of time for an exposure time of one second, with a flux of $3.6\cdot10^6$ electrons/sec. b) Spectral amplitude of the electron fluctuation as a function of the frequency, which corresponds to the Fourier transform of Fig. 2a. c) Acquired count without electron beam as a function of the integration time due to cosmic ray having a flux of 3.1 events/sec from a linear fit. d) Variation of the noise level as a function of the cumulative signal $N$ for different configurations. The experimental curve is fitted with equation \ref{eq:noise_reduced}.}
\end{figure}

We acquired a series of diffraction patterns on a 50 nm thick graphite single crystal. Data were recorded as a function of pump-probe delay time with the shot-to-shot acquisition system. For each acquisition window $(i)$, two diffraction patterns corresponding to $I_{\text{on}}$ and $I_{\text{off}}$ are collected while blocking the photoexcitation pulse before the sample. We sum the electron count over all the pixels to compute the total electron flux. Fig. \ref{fig:fluctuation}a) shows the fluctuation of the normalized scattered electron flux for a single image, $\sum_{\text{px}}I_{\text{on}}$, and for the shot-to-shot acquisition, $\sum_{\text{px}}I_{\text{on}}/\sum_{\text{px}}I_{\text{off}}$. We observe that the fluctuation amplitude is lower by more than one order of magnitude with the shot-to-shot acquisition than with the conventional method. \\
Fig. \ref{fig:fluctuation}b) shows the frequency spectrum of the normalized signal. The shot-to-shot acquisition reduces the noise level at every time regime by a factor of 35 in low frequency and 20 in high frequency.   \\
To quantitatively define the different noise origins, we write the overall noise level as \cite{kealhofer_signal--noise_2015}:
\begin{equation}
    \sigma = \sqrt{\sigma_{\text{shot}}^2+\sigma_{\text{source}}^2+\sigma_{\text{white}}^2+\sigma_{\text{readout}}^2+\sigma_{\text{int}}^2 +\sigma_{\text{gain}}^2 },
    \label{eq:noise}
\end{equation}
where $\sigma_{\text{shot}} = \sqrt{N}$, with $N$ the number of acquired electrons, is the intrinsic shot noise arising from the Poisson-distributed electron beam. \\
The source noise originates from fluctuations in the electron generation and the laser source, which scales as $\sigma_{\text{source}} = \alpha_s \sqrt{f}\sqrt{N}$, with the electron flux $f$ and the RMS magnitude $\alpha_s$. \\
The white noise corresponds to the jitter and thermal effect of the optical and electron pulse generation, scaling as $\sigma_{\text{white}} = \alpha_w N$. \\
The detector noise level, defined as $\sigma_{\text{readout}}$ and $\sigma_{\text{int}}$, depends on the number of counts acquired at instantaneous and different integration times, respectively, when the electron beam is absent. As the single electron detector doesn't have thermal noise like a CCD detector, $\sigma_{\text{readout}} = 0$. This feature is crucial for utilizing the shot-to-shot acquisition method, as it prevents noise accumulation in each acquisition window. The main source of the integration noise $\sigma_{\text{int}}$ is the cosmic rays detected by the single electron detector. However, as shown in Fig. \ref{fig:fluctuation}c), the flux of cosmic rays of 3.1 events/sec is insignificant compared to the electron flux usually set to $\sim 10^6$ electrons/sec. As described in Sec. \ref{sec:part2c}, in the shot-to-shot mode, the exposure window is 25 $\mu s$ per pulse, so over one second, the effective exposure time of the detector is 0.5 s, reducing the integration noise level by a factor of two. The gain noise, $\sigma_{\text{gain}}$, originates from converting a collision event in the silicon detector to a written bit. As explained in Appendix \ref{App:A}, this noise is negligible as long as we don't reach a saturation point of 1 electron/pixel/pulse. \\
The noise level is reduced to 
\begin{equation}
    \sigma = \sqrt{(\alpha \sqrt{N})^2 + (\alpha_w N)^2},
    \label{eq:noise_reduced}
\end{equation}
with $\alpha = (1+\alpha_{s}\sqrt{f})$. Fig. \ref{fig:fluctuation}d) shows the noise level as a function of acquired electrons for single images and shot-to-shot mode for a flux of $f=3.6 \cdot 10^6$ electrons/sec. It shows a source noise of $\alpha_s = 10^{-2}$  Hz$^{-1/2}$ and a white noise of $\alpha_w=1.7\cdot10^{-2}$ for the single image, whereas the noise level of the shot-to-shot acquisition mode reaches the shot noise limit with a source noise of $\alpha_s=1.6\cdot 10^{-4}$  Hz$^{-1/2}$ and a white noise, $\alpha_w = 1.5 \cdot 10^{-16}$, considered negligible. \\
Overall, Fig. \ref{fig:fluctuation} shows that the shot-to-shot mode decreases the noise level by almost a factor of 20. However, it requires twice the integration to build the reference image, so for twice the amount of signal, the SNR still increases by an order of magnitude.\\

\subsection{\label{subsec:3.b} Noise level with pump}
\begin{figure}
\includegraphics[width=\linewidth]{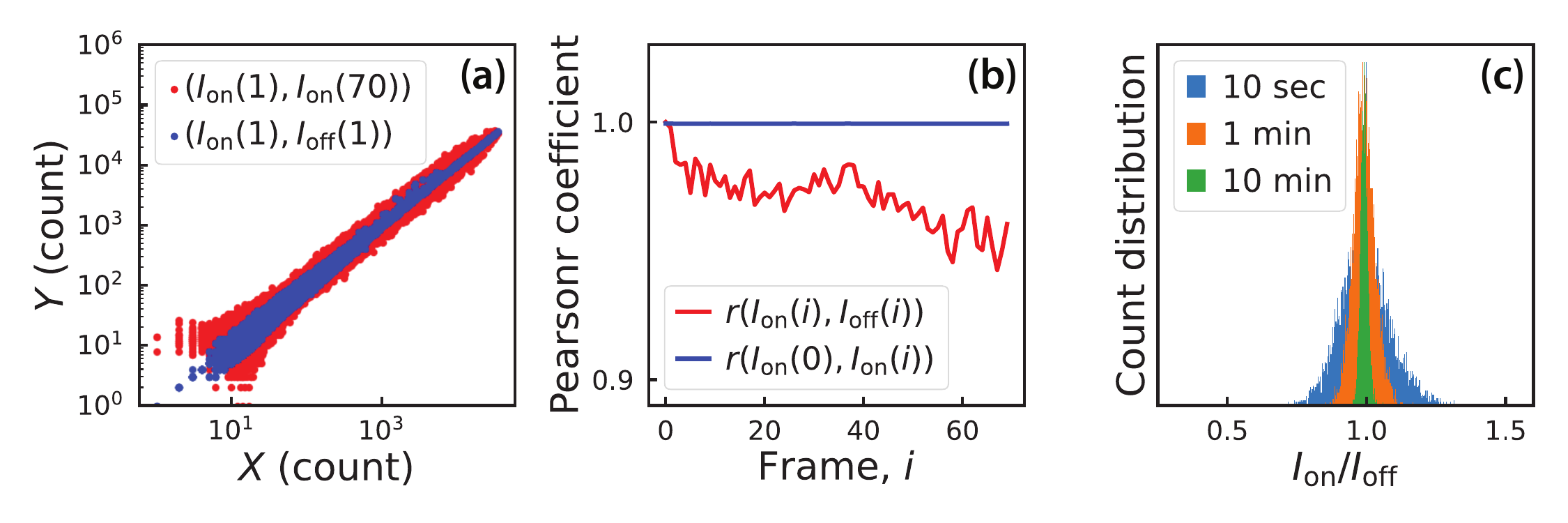}
\caption{\label{fig:fluctuation_pump_on} a) Correlation between pixel values for the two modes of acquisition and b) the Pearson coefficient along frame time. c) Pixel values distribution of $I_{\text{on}}/I_{\text{off}}$ as function of the integration time.}
\end{figure}
In the presence of pump pulses, other noise sources, such as thermal and mechanical fluctuations of the thin film, should be considered. To verify how it affects real measurement, we focus on a pump-probe experiment in which the probe arrives before the pump. The reversible signal after photo-excitation won't interfere with our analysis, but only the long-lived thermal and mechanical fluctuations. \\ 
We take a 50 nm thick sample of natural graphite as a studied specimen whose diffraction patterns are shown in Fig. \ref{fig:real}c). We photoexcite the sample with an absorbed energy density of 0.8 mJ/cm$^2$ at a wavelength around 800 nm, similar to experiments previously reported \cite{carbone_structural_2008, Chatelain2014, stern_mapping_2018}. However, in this section, we analyze the signal before the arrival time of the photoexcitation pulse, when the sample is at equilibrium, which should be identical to the diffraction patterns without photoexcitation. \\
We acquire a set of $I_{\text{on}}^{(i)}$ and $I_{\text{off}}^{(j)}$ images corresponding to a ten-second exposure of graphite's diffraction patterns when it is photoexcited for a time window $i$, and when it is at rest for a time window $j$, respectively. For each diffraction image, we used a mask to set the unscattered electron beam to zero, such that we compare only the fluctuation of the scattered electron. Despite a protection coating on the detector, the pump beam scattered in the chamber creates abnormally high-intensity pixels only visible on the $I_{\text{on}}$ images. To remove them, we compare each image $I_{\text{on}}^{(i)}$ to $I_{\text{off}}^{(i)}$. The pixels in $I_{\text{on}}^{(i)}$ respecting the conditions $I_{\text{on}}^{(i)}/I_{\text{off}}^{(i)}>n$ and $I_{\text{on}}^{(i)}/I_{\text{off}}^{(i)}<1/n$ are set to their value in $I_{\text{off}}^{(i)}$. For a cutting parameter $n=2$, around $1.8\%$ of the pixels are neutralized. \\
Fig. \ref{fig:fluctuation_pump_on}a) shows the correlation between $I_{\text{on}}^{(i)}$ and $I_{\text{off}}^{(i)}$ taken shot-to-shot, and between two diffraction patterns $I_{\text{on}}^{(i)}$ and $I_{\text{on}}^{(j)}$ taken a few seconds apart, that is, $i\neq j$. The pulse-to-pulse images are strongly correlated compared to the two single images, especially at low electron count. We can characterize the correlation between the two images with the Pearson coefficient:
\begin{equation}
    r(I_1, I_2) = \frac{\sum_x (I_1(x) - \langle I_1\rangle)(I_2(x)-\langle I_2 \rangle )}{\sqrt{\sum_x (I_1(x) - \langle I_1\rangle)^2(I_2(x)-\langle I_2 \rangle)^2}},
\end{equation}
where the sum is over all the pixels $x$, and $\langle I \rangle$ corresponds to the mean over the pixels. Fig. \ref{fig:fluctuation_pump_on}b) shows the Pearson coefficient for a set of integrated images taken at different times. The correlation between $I_{\text{on}}^{(i)}$ and $I_{\text{off}}^{(i)}$ images taken shot-to-shot stays close to 1, while the one between $I_{\text{on}}^{(0)}$ and $I_{\text{on}}^{(j)}$ decreases and fluctuates with the time between exposure shots. The average correlation between $\langle I_{\text{on}}(i), I_{\text{off}}(i) \rangle_i$ is $1-6.8\cdot 10^{-4}$, with a standard deviation of $1.6\cdot10^{-5}$. It suggests that the shot-to-shot acquisition mode effectively removes thermal and mechanical fluctuations. \\
Fig. \ref{fig:fluctuation_pump_on}c) shows the pixel distribution of the normalization $I_{\text{on}}/I_{\text{off}}$ as a function of exposure time. We observe that it converges towards a Gaussian distribution with decreasing variance, reaching $0.01$ with a mean of $0.99$. It confirms the robustness of the correlation between the diffraction pattern of the unpumped and pumped graphite. 

\section{\label{sec:4} Realistic situation}
The analysis of the noise origins showed that the SNR increased by one order of magnitude for the shot-to-shot acquisition with respect to the conventional technique. The correlation between the pumped and the unpumped data also showed that this acquisition method negates mechanical fluctuations in a pulse-to-pulse regime. We now focus on time traces after photoexcitation for two different practical cases: high-fluence pump on a copper grid generating a cloud of free electrons, which forms an electric field deviating the electron beam; and a well-studied sample, graphite, perturbed with a low fluence. 

\subsection{Plasma generation}

\begin{figure}
\includegraphics[width=0.7\linewidth]{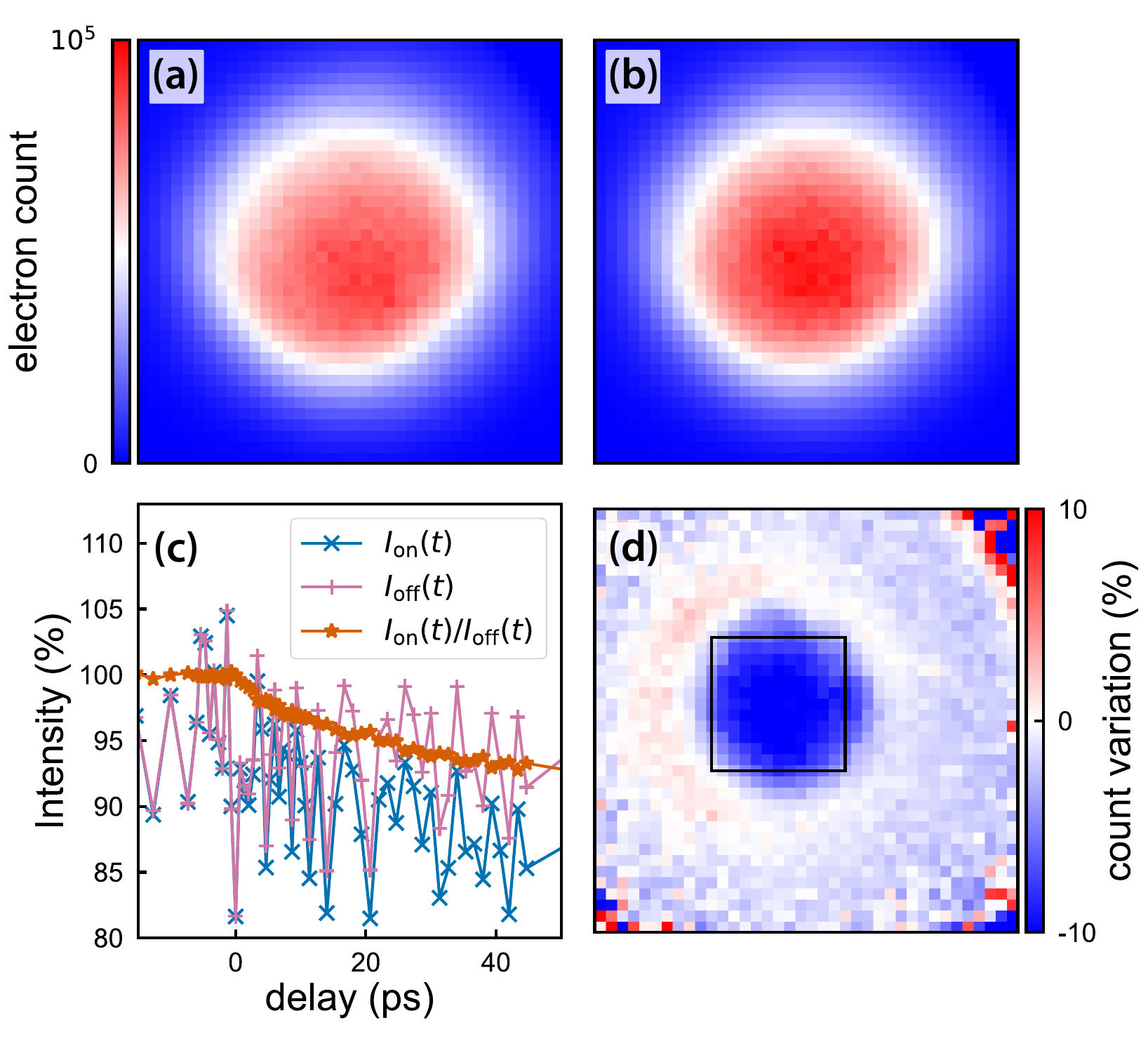}
\caption{\label{fig:plasma} Transient signal due to plasma generation after photoexcitation on a copper grid. Zoom in on the unfocused electron beam that went through the copper grid after the photoexcitation photon pulse in a) the presence of the pump $I_{\text{on}}$, b) the absence of the pump $I_{\text{off}}$, and d) the normalization between each other $I_{\text{on}}/I_{\text{off}}$. c) Comparison between the transient signals measured for a region of interest in the center of the electron beam $I_{\text{on}}(t)$, $I_{\text{off}}(t)$, and $I_{\text{on}}(t)/I_{\text{off}}(t)$.}
\end{figure}

We use a 1000 mesh copper grid and photoexcite at a wavelength of 800 nm with a fluence of 2 mJ/cm$^2$. Fig. \ref{fig:plasma}a) and b) show the unfocused electron beam that went through the copper grid in the absence, $I_{\text{on}}(t)$, and the presence, $I_{\text{off}}(t)$, of the pump, respectively, at time $t$ after the photoexcitation. Thanks to the shot-to-shot acquisition detection, the normalization between Fig. \ref{fig:plasma}a) and b), $I_{\text{on}}/I_{\text{off}}$, shown in Fig. \ref{fig:plasma}d), clearly reveals the pattern of the plasma after the photoexcitation.\\
Fig. \ref{fig:plasma}c) shows the total count in a region of interest defined as a $6\times6$ pixels highlighted by the square in Fig. \ref{fig:plasma}d). Each delay corresponds to an exposure time $20\cdot10^3$ pulses for both  $I_{\text{on}}(t)$, and $I_{\text{off}}(t)$. No transient signal can be observed for the single image acquisition since the noise level of $>10\%$ overcomes the physical signal of $5\%$. However, the normalization $I_{\text{on}}(t)/I_{\text{off}}(t)$ for each delay $t$ increases the SNR by almost two orders of magnitude, which reveals the physical transient signal due to the deflection of the electrons by the plasma generated after the photoexcitation.\\

\subsection{Graphite}

\begin{figure}
\includegraphics[width=\linewidth]{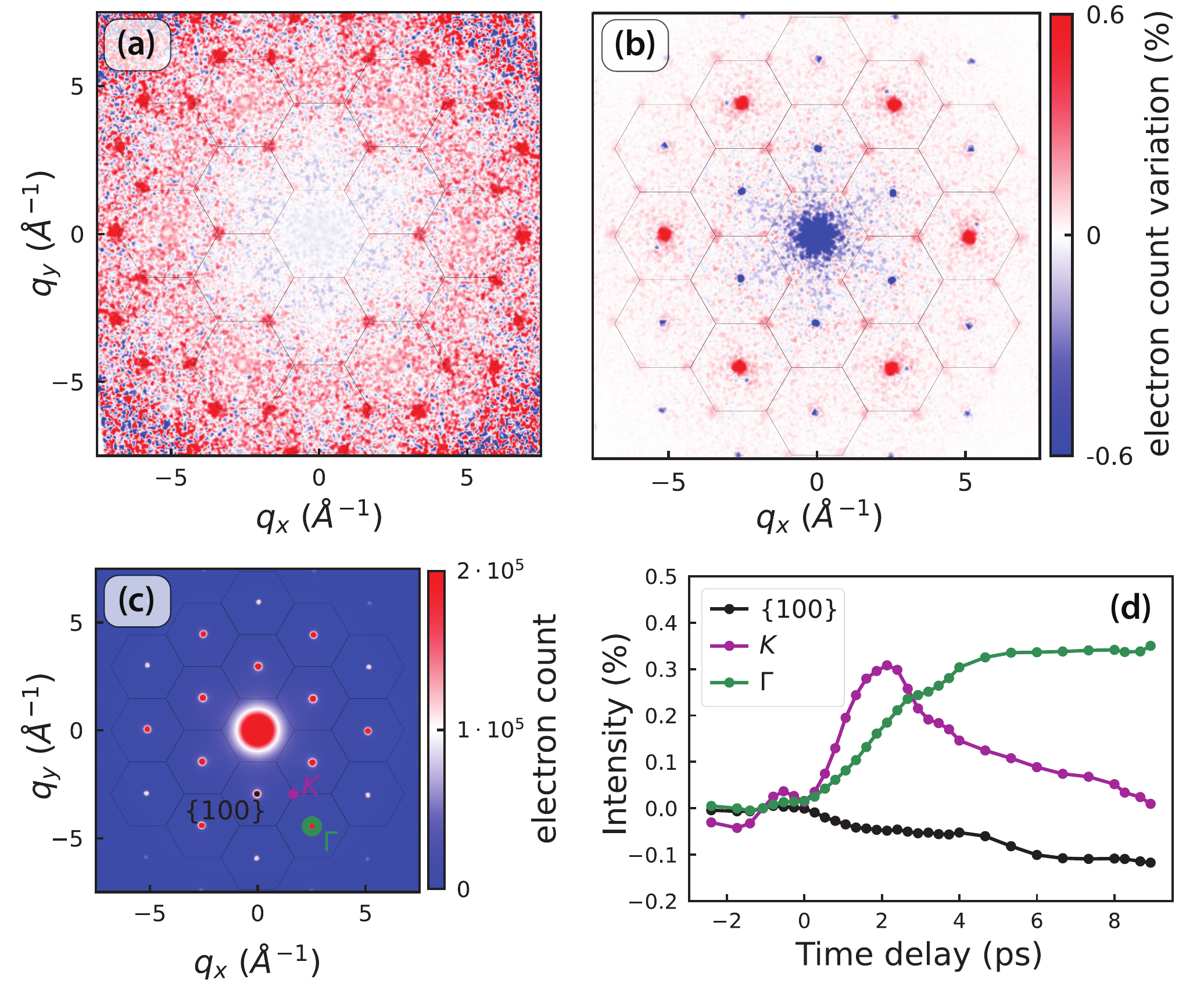}
\caption{\label{fig:real} a) and b) are the normalized diffraction patterns as equation \ref{eq:deltaI} and \ref{eq:deltaI-} respectively, averaged over a time delay, $t$, between 1 and 2 ps. c) 6-fold averaged diffraction patterns of graphite, corresponding to $I_{\text{on}}(t=-2.4 \text{ ps}, \boldsymbol{q})$. d) Electron count variation for specific regions of interest highlighted in panel c) corresponding to the Bragg peak family ${100}$ and the diffuse scattering at reduced momentum $K$ and around the $\Gamma$ point. }
\end{figure}

The photo-induced plasma generation already shows a striking difference between the conventional and shot-to-shot acquisition methods. However, the intensity distribution with diffraction patterns relies on the sample geometry. Photo-induced thermal and mechanical effects can induce instability or artifacts. In Sec. \ref{subsec:3.b} we already showed that these effects are stable over time. In this section, we confirm the capabilities of the shot-to-shot acquisition method on a time-resolved experiment on graphite. \\
Ultrafast electron diffraction on graphite has already been extensively studied \cite{carbone_structural_2008, chatelain_ultrafast_2012, stern_mapping_2018}. To account for the sensitivity of our acquisition techniques, we use a low absorbed excitation energy of 0.8 mJ/cm$^2$ on a 50 nm thick flake of natural graphite. The diffraction patterns are acquired over 24 hours, for an exposure of 22 min per delay (11 min for $I_{\text{on}}(t)$ and $I_{\text{off}}(t)$ respectively). To overcome the slight zone axis misalignment and gain SNR, we averaged the diffraction patterns along the 6-fold symmetry axis. \\
To observe the photo-induced effect, we normalize the excited diffraction patterns by the one at rest and subtract this by the same ratio averaged along delays before time zero, $t_0$. This subtraction removes the constant thermal effect. It can be written as: 
\begin{equation}
    \Delta I (t, \boldsymbol{q}) = \frac{I_{\text{on}}(t,\boldsymbol{q})}{I_{\text{off}}(t,\boldsymbol{q})}-\frac{\langle I_{\text{on}}(t,\boldsymbol{q})\rangle_{t<t_0}}{\langle I_{\text{off}}(t,\boldsymbol{q}) \rangle_{t<t_0}} \label{eq:deltaI}
\end{equation}
This normalization is shown in Fig. \ref{fig:real}a) at a time delay $t=1$ ps after the photoexcitation. Since the normalization occurs pixel by pixel, the noise increases significantly in low-intensity regions. This effect is visible on the side of the figure where the signal is the lowest. To avoid this effect, we can subtract the pumped diffraction pattern from the unpumped as: 
\begin{multline}
    \delta I (t, \boldsymbol{q}) = \frac{I_{\text{on}}(t,\boldsymbol{q})}{\langle I_{\text{on}}(t,\boldsymbol{q})\rangle_{\boldsymbol{q}}} - \frac{I_{\text{off}}(t,\boldsymbol{q})}{\langle I_{\text{off}}(t,\boldsymbol{q})\rangle_{\boldsymbol{q}}} - \biggl< \frac{I_{\text{on}}(t,\boldsymbol{q})}{\langle I_{\text{on}}(t,\boldsymbol{q})\rangle_{\boldsymbol{q}}} - \frac{I_{\text{off}}(t,\boldsymbol{q})}{\langle I_{\text{off}}(t,\boldsymbol{q})\rangle_{\boldsymbol{q}}} \biggr>_{t<t_0}
    \label{eq:deltaI-}
\end{multline}
Fig. \ref{fig:real}c) shows this subtraction for a time delay $t=1$ ps. In this case, there is a strong signature where the electron intensity is large, at the Bragg peak position, but we can still resolve the low variation in the diffuse scattering at the K-point. \\
The transient dynamics account for the sensitivity of the setup qualitatively. Fig. \ref{fig:real}d) shows the intensity variation for the ${100}$ Bragg peak family and the diffuse scattering at the $K$-point and around the $\Gamma_{{110}}$ Bragg peak highlighted in the diffraction pattern Fig \ref{fig:real}c). The low excitation energy induces a diffraction decrease of only $0.1\%$ due to the Debye-Waller effect. We can still resolve the transition between the fast initial drop from the generation of strongly coupled optical (SCOP) phonon to the slower one from their decay to acoustic phonon at 1 ps. This is consistent with the rapid increase of the diffuse scattering intensity at the K-point due to the generation of $A_1'$ optical phonon. Its population starts decreasing at 2 ps as it scatters with lower energy near $\Gamma$ momentum acoustic phonons as suggested by the transient dynamics of diffuse scattering around the $\Gamma_{110}$ Bragg peak family. These results are consistent with the previous work of Stern \textit{et al} \cite{stern_mapping_2018} in which the photoexcitation energy is 15 times higher than the one used here. \\

\section{Conclusion}

Similar to ultrafast optical spectroscopy, the shot-to-shot acquisition method significantly improves the signal-to-noise ratio for the study of transient dynamics in solids and molecules. By phase locking a high acquisition rate direct electron detector, the noise reaches the shot noise limit. By introducing a photoexcitation pulse, we observe thermal and mechanical fluctuations that are stable over time and don't introduce additional white noise. \\
The shot-to-shot acquisition technique revealed a marked improvement over conventional methods. The investigation of plasma generation and transient dynamics in graphite diffraction patterns demonstrates a time resolution of approximately 1 ps, achieved with an electron flux of just 1000 electrons per pulse. In the absence of a pulse compression scheme, the temporal resolution varies with the electron flux, improving at lower flux and degrading as the flux increases. \\
The high sensitivity of the setup enables the investigation of ultrafast lattice dynamics following low-level photoexcitation. In this regime, the lattice temperature increases by only a few tens of degrees, allowing the observation of dynamics driven by fluctuations in electronic, magnetic, or collective excitations.\\
This acquisition technique offers sensitivity comparable to, or exceeding, that of other UED configurations operating at high repetition rates \cite{freelon_design_2023} or using high electron flux \cite{chatelain_ultrafast_2012}, while providing enhanced sensitivity in the low-perturbation regime. Integrating this approach with a higher repetition-rate laser would increase the signal by one to two orders of magnitude, without compromising temporal resolution or beam quality.\\
Shot-to-shot acquisition schemes can be implemented in other electronic spectroscopy methods, such as ultrafast transmission electron microscopy (UTEM)\cite{Zewail_4D}, ultrafast Lorentz TEM\cite{tengdin_imaging_2022}, time-resolved electron energy loss spectroscopy (tr-EELS)\cite{carbone_dynamics_2009}, time- and angle-resolved photoemission spectroscopy (tr-ARPES)\cite{na_direct_2019}, and others. This paves the way for investigating sensitive materials in the low-perturbation regime.

\begin{acknowledgments}
We acknowledge the useful discussion with the Dectris team, especially Jan V\'avra, who helped us install the detector.\\

We acknowledge support from the SNF grant 200331. B.W. acknowledges support from the EU H2020 research and innovation program under the Marie Skłodowska-Curie Actions grant agreement 801459 (FP-RESOMUS). This project has also received funding from the European Union's Horizon 2020 research and innovation program under grant agreement no. 871124 Laserlab-Europe.
\end{acknowledgments}

\section*{Data Availability Statement}
The data that support the findings of
this study are available from the
corresponding author upon reasonable
request. 

\appendix
\section{\label{App:A}Quantum detection efficiency}

\begin{figure}
    \centering
    \includegraphics[width=\linewidth]{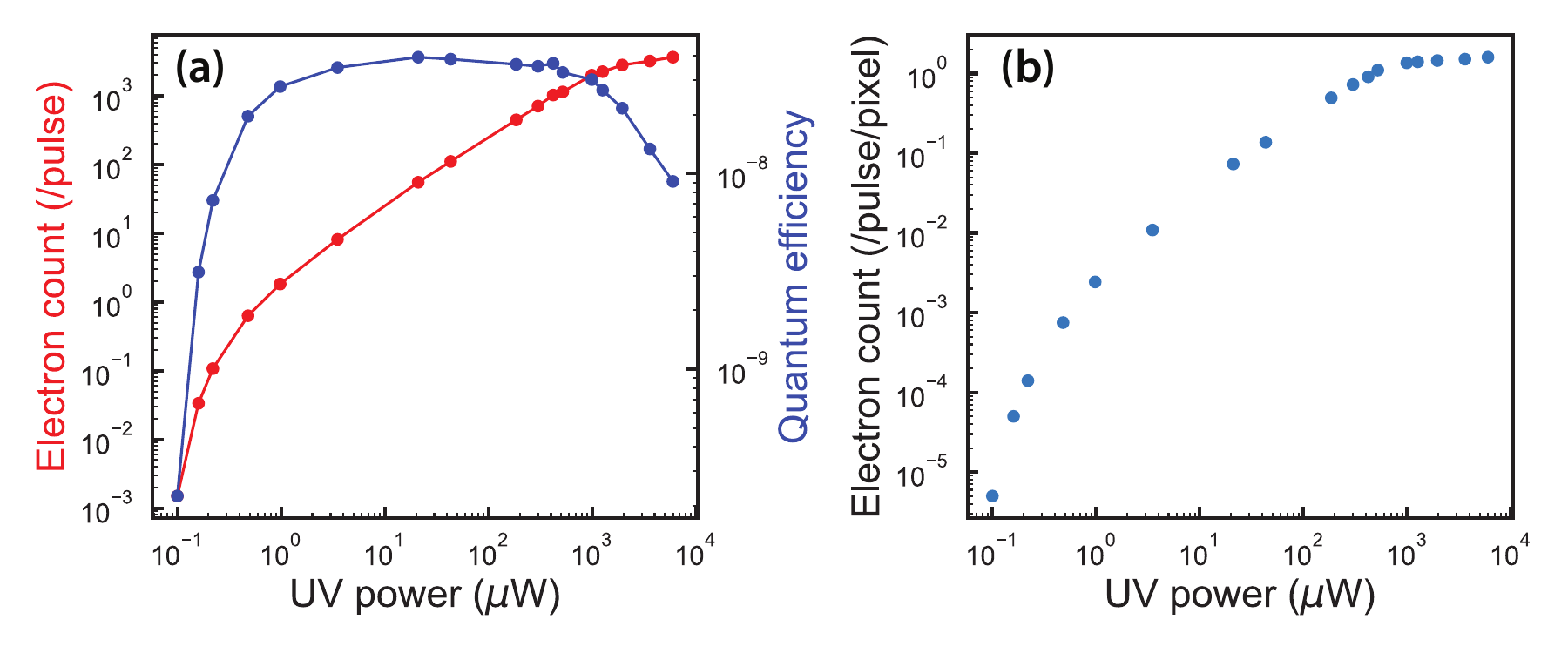}
    \caption{(a) Electron flux and quantum efficiency (Eq. \ref{eq:QDE}) as a function of the input UV power for an electron energy of 30 keV. (b) Acquired electrons per pulse for one selected pixel as a function of the UV power.}
    \label{fig:qde}
\end{figure}

When an electron collides with a detector's pixel, charges are excited, and a current is measured. If it reaches a given threshold, the detector counts one electron. The threshold is factory-calibrated for different electron energies.\\ 
We are interested in finding the optimal working conditions of the electron beam. One way to characterize the electron beam is the quantum efficiency, defined as
\begin{equation}
    QE = \frac{n_{\text{el}}}{n_{\text{ph}}},
    \label{eq:QDE}
\end{equation}
where $n_{\text{el}}$ is the number of acquired electrons per pulse directly measured by the direct electron detector for one-second exposure time. $n_{\text{ph}}=P/fE_{\text{ph}}$ is the number of incident photons per pulse for a repetition rate $f=20$ kHz, a photon energy $E_{\text{ph}}=4.66$ eV and beam power $P$ measured with a power meter.\\
Fig. \ref{fig:qde}a) shows the acquired electron, $n_{\text{el}}$ and the quantum efficiency, $QE$, as a function of the input UV power. We observe three stages with a non-linear increase of the $QE$ at $P<1 \mu$W, a plateau between 1 $\mu$W and 1 mW, and a decrease of the $QE$ above 1 mW. The first stage corresponds to an extremely low dose rate, below one electron per pulse, so undetectable by the detector, while the last stage corresponds to a flux saturation. The latter can arise from a saturation in the photoemission due to the space charge and thermal effects of the photocathode, the Child-Langmuir limit, or from a saturation of the detection. The detector counts one electron for a given pixel when its signal reaches a threshold. When the signal is higher than the threshold for longer than 200 ns, the detector counts another electron. This retrigger functionality enables high electron flux measurement for continuous beam, but for a short pulsed beam of time duration $\ll$~200~ns, the electron bunch is too dense, such that the detector cannot acquire more than one electron per pixel per pulse. Fig. \ref{fig:qde}b) illustrates this by showing the electron count for one pixel as a function of the input UV power. From one electron/pulse/pixel, the acquired count starts saturating. It is not favored to use a high pulsed electron flux with this kind of direct electron detector, as artifacts can arise from the detection's saturation. \\
As shown in Fig. \ref{fig:qde}, the optimal UV power range for electron photoemission lies between 1 $\mu$W and 1 mW, where the measured quantum efficiency reaches a maximum of approximately $4\cdot10^{-8}$ electrons per photon. \\

\section{\label{App:B}Electronic response of the detector to an electron pulse}
\begin{figure}
    \centering
    \includegraphics[width=0.7\linewidth]{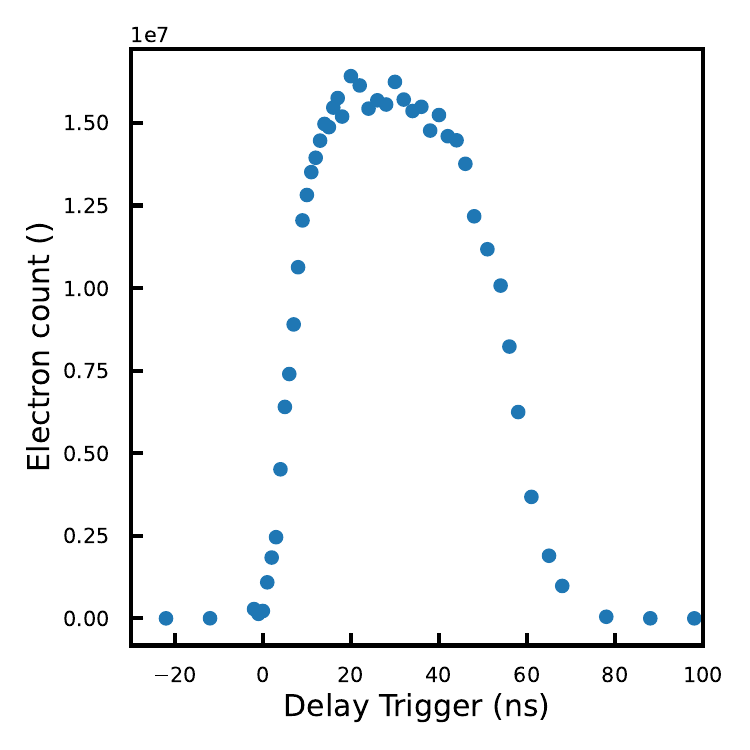}
    \caption{Electron count measured by the Quadro as a function of the delay given to a square signal with fixed width.}
    \label{fig:recovery}
\end{figure}

Electron detectors can have a long recovery time after an electron bunch arrival. If the recovery time is longer than the delay between two pulses, the shot-to-shot acquisition can lose its benefits. \\
To quantify the electronic response of the direct electron detector we vary the delay of the acquisition window from the laser trigger with a fixed width of 10 ns. The delay can be tuned with steps down to 1 ns. For each delay, we acquire an image of the raw electron beam with one-second exposure in shot-to-shot acquisition mode. Fig. \ref{fig:recovery} shows the total electron count as a function of the trigger delay. Even though the electron pulse duration is around 1 ps, the signal acquired by the detector lasts 80 ns. Then, the recovery time is smaller than the one between the pulse ($50 \mu$s). 

\newpage

\nocite{*}
\bibliography{main}
\end{document}